\documentclass[twocolumn]{aastex631}

\newcommand{\Mjup}{\mbox{$\rm M_\mathrm{Jup}$}}
\newcommand{\Msun}{\mbox{$\rm M_{\odot}$}}
\newcommand{\Rsun}{\mbox{$\rm R_{\odot}$}}

\newcommand{\e}[1]{\times 10^{#1}}

\shorttitle{$\delta$ Scuti Pulsations and plausible Spin-Orbit Alignment for HIP~65426}
\shortauthors{Sepulveda et al.}

\begin{document}

\title{HIP~65426 is a High-Frequency Delta Scuti Pulsator in Plausible Spin-Orbit Alignment with its Directly Imaged Exoplanet}

\correspondingauthor{Aldo G. Sepulveda}
\email{aldo.sepulveda@hawaii.edu}
\author[0000-0002-8621-2682]{Aldo G. Sepulveda}
\altaffiliation{NSF Graduate Research Fellow}
\affiliation{Institute for Astronomy, University of Hawai`i at M\={a}noa. 2680 Woodlawn Drive, Honolulu, HI 96822, USA}
\author[0000-0001-8832-4488]{Daniel Huber}
\affiliation{Institute for Astronomy, University of Hawai`i at M\={a}noa. 2680 Woodlawn Drive, Honolulu, HI 96822, USA}
\affiliation{Sydney Institute for Astronomy (SIfA), School of Physics, University of Sydney, NSW 2006, Australia}
\author[0000-0001-5222-4661]{Timothy R. Bedding}
\affiliation{Sydney Institute for Astronomy (SIfA), School of Physics, University of Sydney, NSW 2006, Australia}

\author[0000-0003-3244-5357]{Daniel R. Hey}
\affiliation{Institute for Astronomy, University of Hawai`i at M\={a}noa. 2680 Woodlawn Drive, Honolulu, HI 96822, USA}

\author[0000-0002-5648-3107]{Simon J. Murphy}
\affiliation{Centre for Astrophysics, University of Southern Queensland, Toowoomba, QLD 4350, Australia}

\author[0000-0002-3726-4881]{Zhoujian Zhang}
\altaffiliation{NASA Sagan Fellow}
\affiliation{Department of Astronomy and Astrophysics, University of California, Santa Cruz, CA 95064, USA}

\author[0000-0003-2232-7664]{Michael C. Liu}
\affiliation{Institute for Astronomy, University of Hawai`i at M\={a}noa. 2680 Woodlawn Drive, Honolulu, HI 96822, USA}

\submitjournal{The Astronomical Journal}
\accepted{May 07, 2024}

\begin{abstract}
HIP~65426 hosts a young giant planet that has become the first exoplanet directly imaged with JWST. Using time-series photometry from the Transiting Exoplanet Survey Satellite (TESS), we classify HIP~65426 as a high-frequency $\delta$~Scuti pulsator with a possible large frequency separation of $\Delta \nu =$7.23$\pm$0.02~cycles~day$^{-1}$. We check the TESS data for pulsation timing variations and use the nondetection to estimate a 95\% dynamical mass upper limit of 12.8~\Mjup\ for HIP~65426~b. We also identify a low-frequency region of signal that we interpret as stellar latitudinal differential rotation with two rapid periods of 7.85$\pm$0.08~hr and 6.67$\pm$0.04~hr. We use our TESS rotation periods together with published values of radius and $v \sin{i}$ to jointly measure the inclination of HIP~65426 to $i_{\star}=107_{-11}^{+12}$$^\circ$. Our stellar inclination is consistent with the orbital inclination of HIP~65426~b ($108_{-3}^{+6}$$^{\circ}$) at the $68\%$ percent level based on our orbit fit using published relative astrometry. The lack of significant evidence for spin-orbit misalignment in the HIP~65426 system supports an emerging trend consistent with preferential alignment between imaged long-period giant planets and their host stars.
\end{abstract}


\section{Introduction} \label{sec:intro}
HIP~65426 (HD~116434, TIC~438702139) is a nearby (107.5$\pm$0.4~pc, \citealt{gaiamission,GaiaEDR3}) A2~V star \citep{Houk1978} that hosts a directly imaged giant planet \citep[HIP~65426~b,][]{Chauvin+2017}. The host star is a member of the Lower Centaurus-Crux (LCC) young moving group \citep{deZeeuw+1999,Luhman2022}, with \textbf{an} age of $\sim$10--23~Myr \citep[e.g.,][]{Mamajek+2002,Sartori+2003,Pecaut+2012,Song+2012}. Independent of its membership in LCC, \citet{Chauvin+2017} derived an age of 14$\pm$4~Myr for HIP~65426 primarily using isochrone fitting. HIP~65426 has several astrophysically interesting qualities in addition to being an exoplanet host star. Its $v \sin{i}$ (261--299~km~s$^{-1}$, \citealt{Chauvin+2017,Petrus+2021}) indicates that it is rapidly rotating. Hints of stellar pulsations have been identified in its radial velocity variations \citep[e.g.,][]{Chauvin+2017,Grandjean+2020,Petrus+2021}, but HIP~65426's variability status has not yet been formally confirmed and classified. A detection of high-frequency $\delta$~Scuti pulsations would independently confirm HIP~65426's youth \citep{Bedding+2020Natur} and might even enable an asteroseismic age to be measured \citep[e.g.,][]{Murphy+2021,Steindl+2022}.

The giant planet HIP~65426~b orbits at a semi-major axis of 62--120 au and inclined 99--112$^{\circ}$ to the line of sight based on orbit fitting analyses  \citep{Cheetham+2019,Bowler+2020,Carter+2023,DoO+2023,Blunt+2023}. Mass estimates of HIP~65426~b, derived from evolution models and from spectral fitting, span 6--11 \Mjup\ \citep{Chauvin+2017,Cheetham+2019,Marleau+2019,Petrus+2021,Carter+2023}. HIP~65426 was recently observed with NIRCam and MIRI as part of the JWST \citep{Gardner+2006,Gardner+2023} Early Release Science Program \citep[][]{Hinkley+2022}, resulting in HIP~65426~b becoming the first direct detection of an exoplanet at wavelengths beyond 5~$\mu$m \citep{Carter+2023}.

Assessment of the spin-orbit alignment between a host star and its orbiting companions (i.e., obliquity) provides insight into the dynamical history of the system. While obliquity measurements have been obtained for dozens of short-period transiting exoplanets \citep[e.g.,][]{Albrecht+2022}, there are relatively few constraints for long-period imaged substellar companions \citep[e.g.,][]{Bowler+2017,Bryan+2020,Kraus+2020}. Recently, \citet{Bowler+2023} analyzed the spin-orbit alignments for a sample of 23 imaged substellar companions orbiting cool stars by constraining the stellar inclinations and comparing to the companion orbital inclinations. They found that misalignments are common for their sample, which mostly comprised brown dwarfs. However, the two imaged giant planet systems in their sample were consistent with alignment or near-alignment. Determining whether or not additional imaged giant planets like HIP~65426~b are aligned with their host star would contribute to our understanding of giant planet formation as a population, as well as help to clarify the dynamical history for the planet itself.  

Time-series photometry from the Transiting Exoplanet Survey Satellite \citep[TESS,][]{TESSMission} is continuing to boost the sample size of systems where obliquity assessments can be carried out. Time-series photometry can reveal the stellar rotation period if surface spot modulation is significant \citep[e.g.,][]{Affer2012,McQuillan+2014}. While this has traditionally been carried out for cool stars, \citet{Balona2011,Balona2013,Balona2017} used Kepler time-series photometry to show that as many as $\sim$40\% of A-type stars exhibit a detectable photometric rotation frequency. This fraction persists to $\sim$30\% when source contamination is considered \citep{Sikora+2020}. 
A measured rotation period places a determination on the line-of-sight inclination when combined with the stellar radius and spectroscopic projected rotational velocity ($v \sin{i}$), \citep[e.g.,][]{Masuda+Winn2020}. Furthermore, time-series photometry also probes for other phenomena, including stellar pulsations and transit events, both of which have previously been studied in directly imaged exoplanet host stars using TESS \citep[e.g.,][]{Zieba+2019,Pavlenko+2022,Sepulveda+2022,Sepulveda+2023}. The challenge for hot A-stars like HIP~65426 then becomes a matter of detecting a rotation frequency and discerning it from other phenomena, such as pulsation modes or contaminating sources in the aperture.

We analyze 2-minute cadence TESS time-series photometry of HIP~65426 in this study. We present the detection of stellar pulsations and rotational modulation in the data, and use the latter to constrain the obliquity of the system. We conclude by summarizing the implications of our analysis.

\section{TESS Observations}\label{sec:obs}
HIP~65426 was observed by TESS in 2-minute cadence mode for Sectors 11, 38, and 64. The gaps between these sectors are $\sim$2~yr. Sector 11 observations spanned UTC 23 April 2019--20 May 2019, Sector 38 spanned UTC 29 April 2021--26 May 2021, and Sector 64 spanned UTC 06 April 2023--04 May 2023. We used \texttt{lightkurve} \citep{LightkurveCollaboration+2018} to download PDC-SAP light curves processed by SPOC \citep{smith12,stumpe12,stumpe14,jenkins16}. We removed outlying photometry from each TESS sector via a sigma-clipping of five standard deviations from the median flux value, which is a standard practice to remove outliers potentially caused by instrumental artifacts that were not captured through standard quality flags. This resulted in only 21 cadences ($<0.05\%$) being removed, resulting in 16,392 cadences for Sector 11, 18,486 cadences for Sector 38, and 18,849 cadences for Sector 64.

We assessed our TESS data for any significant sources of contamination. We began by inspecting a region within a radius of 80\arcsec\ from HIP~65426, which is an approximate size of the default aperture used for our light curves. We identified no Gaia sources with comparable brightness to HIP~65426, which has an apparent Gaia $G$-band magnitude of 7.00 \citep{gaiamission,GaiaDR3}. The five brightest Gaia sources after HIP~65426 in this region (spanning separations of $\sim$17--47\arcsec) have respective apparent $G$-band magnitudes of 11.98, 13.28, 14.12, 14.59, and 15.05. Their respective $G_{\rm BP}-G_{\rm RP}$ colors are 1.72, 0.72, 1.41, 1.47, and 1.32. HIP~65426 has no known stellar companions \citep{Kouwenhoven+2005,Chauvin+2017,Petrus+2021}. Furthermore, the TESS Input Catalog (TIC, \citealt{Stassun+2018,Stassun+2019}) gives a flux contamination ratio (i.e., ratio of the total contaminant flux within 210\arcsec\ to HIP~65426's flux) of only 0.03 (See also Section \ref{sec:rot}). Overall, it is likely that significant signals in our TESS data correspond to HIP~65426.
\vspace{1mm}
\begin{figure*}[h!]
  \includegraphics[trim=0cm 0cm 0cm 0cm, clip, width=7.2in]{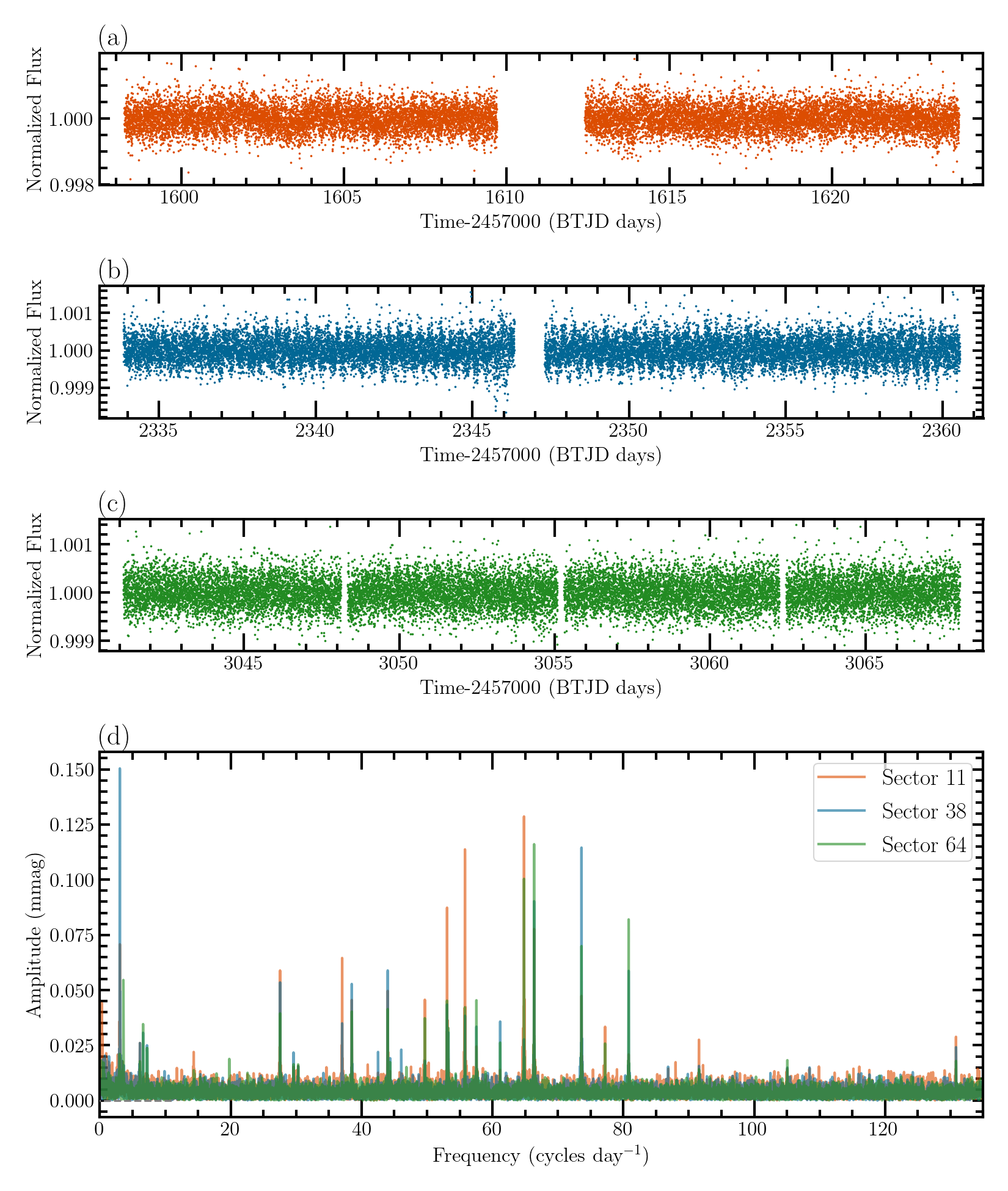}
  \centering
  \caption{(a--c): TESS time-series photometry of HIP~65426 for Sectors 11, 38, and 64. (d): Amplitude spectra of each TESS sector in linear-linear space.
  \label{fig:TESSobs}} 
\end{figure*}

\begin{figure*}
  \includegraphics[trim=0cm 0cm 0cm 0cm, clip,width=7.1in]{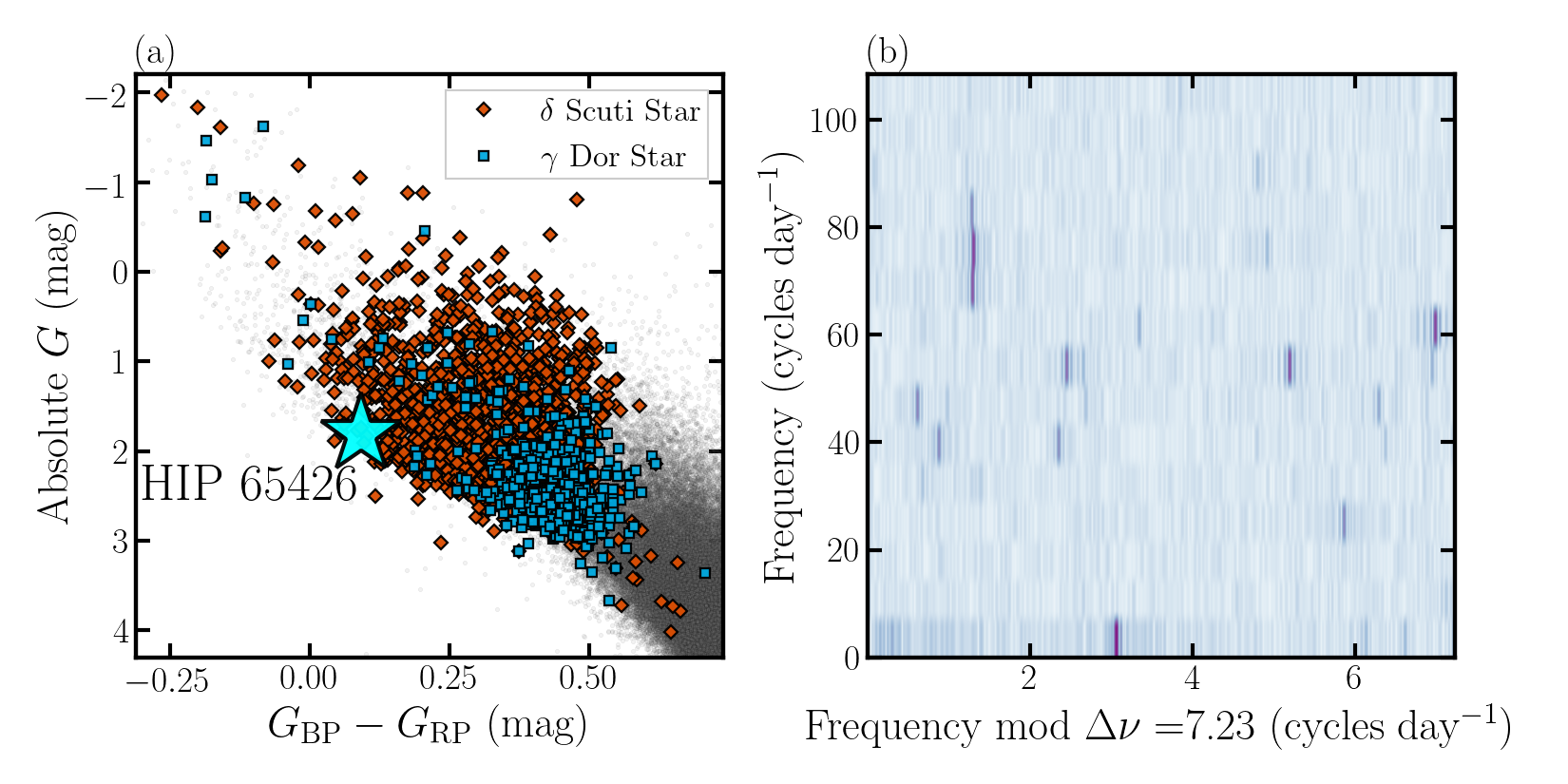}
  \centering
  \caption{(a) Gaia color-magnitude diagram for a sample of Kepler stars generated following \citet{Sepulveda+2022}. HIP~65426 is overplotted as a large cyan star. The orange diamonds are the $\delta$~Scuti stars, blue squares are the $\gamma$~Dor stars, and the small, grey, semi-transparent dots are the general population of Kepler stars. (b): \'{e}chelle diagram for HIP~65426 generated using our suggested large frequency separation of $\Delta \nu=$7.23 cycles~day$^{-1}$. This plot uses an amplitude spectrum calculated from the concatenated TESS time series. \label{fig:deltascuti}} 
\end{figure*}
\section{Analysis}
\subsection{High-Frequency $\delta$~Scuti Pulsations in HIP~65426}\label{sec:dspulsation}
We calculated the amplitude spectrum of each TESS sector time series (Figure \ref{fig:TESSobs}abc) using \texttt{lightkurve} \citep{LightkurveCollaboration+2018} and identified several significant frequencies spanning  $\sim$3--130 cycles~day$^{-1}$ (Figure \ref{fig:TESSobs}d). For the high frequencies, we used \texttt{SigSpec} \citep{SigSpec} to iteratively fit sine waves to the data above 10~cycles~day$^{-1}$, and up to a spectral significance threshold of 10 \citep[following, e.g.,][]{2007MNRAS.379.1498G, 2009A&A...494.1031Z, Sepulveda+2022}. This resulted in 13 significant pulsation frequencies for Sector 11 and 14 significant pulsation frequencies for Sectors 38 and 64. The pulsation frequencies span 28--131~cycles~day$^{-1}$, consistent with pressure modes of $\delta$~Scuti stars \citep[e.g.,][]{Kurtz2022}, and they are tabulated in Table \ref{tab:freqs}.

Figure \ref{fig:deltascuti}a shows HIP~65426 in a Gaia DR3 \citep[][]{gaiamission,GaiaDR3} color-magnitude diagram (CMD) compared to the Kepler $\delta$~Scuti stars from \citet{Murphy+2019} as well as the Kepler $\gamma$~Doradus stars from \citet{Li+2020}. We created the CMD following the procedure described in \citet{Sepulveda+2022}, which includes corrections for extinction and reddening for the Kepler stars. We used $A_V = 0.038$ mag \citep{Chen+2012} for HIP~65426. The $G_{\rm BP}-G_{\rm RP}$ color and absolute $G$ magnitude of HIP~65426 are consistent with the population of $\delta$~Scuti stars from the Kepler Mission (Figure \ref{fig:deltascuti}a) as well as with the population of high-frequency $\delta$~Scuti stars recently detected in the Pleiades open cluster with TESS \citep{Bedding+2023}.

We attempted an asteroseismic mode identification for HIP~65426 by constraining the large frequency separation ($\Delta \nu$), which is a parameter that can also constrain the mean stellar density \citep{Aerts+2010}. We used the \texttt{echelle} package \citep{Hey+Ball2020} to iterate through a range of trial $\Delta \nu$ values (5--9~cycles~day$^{-1}$, which generously encompasses typical $\Delta \nu$ values, e.g., \citealt{Bedding+2020Natur,Murphy+2023}) and visually inspected the \'{e}chelle  diagrams for regular vertical sequences following \citet{Bedding+2020Natur}. We arrived at a possible value of $\Delta \nu=$7.23$\pm$0.02~cycles~day$^{-1}$ (Figure \ref{fig:deltascuti}b). This is consistent within the range that \citet{Bedding+2020Natur} found for their sample of high-frequency $\delta$~Scuti stars, but the resulting ridge patterns of HIP 65426 are too complex to confidently identify the majority of the pulsation modes. The rapid stellar rotation period (indicated by $v \sin{i}$ and our analysis in Section \ref{sec:rot}) greatly contributes to the complication \citep[e.g.,][and references therein]{Reese2022,Aerts+Tkachenko2023}. Detailed asteroseismic modeling, which may yield an independent age estimate, is beyond the scope of this study.    

We checked the TESS data for pulsation timing variations that could be caused by mutual gravitation with an orbital companion \citep[e.g.,][]{Hey+2020}. Such effects would manifest as a periodic (or coherent) variation of the phase of all the $\delta$ Scuti pressure modes, and is only measurable for sufficiently massive planets on long enough periods \citep[e.g.,][]{Murphy+2016,Hey+2021}. However, if no variation is detected, it is possible to instead place an upper limit on the mass of the orbital companion known a priori to exist. This is also a useful tool to rule out stellar mass companions. We follow the same methodology as \citet{Hey+2021} to place a dynamical mass upper limit on HIP~65426~b in the absence of coherent phase variations, using our posterior predictions of the orbit from Section \ref{sec:orbFit}. This results in a 95\% upper limit of 12.8~\Mjup, which serves as an independent verification that HIP~65426~b is likely in the giant planet mass regime. The 99.7\% upper limit is 19.1~\Mjup.

\subsection{Rapid Differential Rotation in HIP~65426}\label{sec:rot}
\begin{figure*}[h!]
  \includegraphics[trim=0cm 0cm 0cm 0cm, clip, width=7.3in]{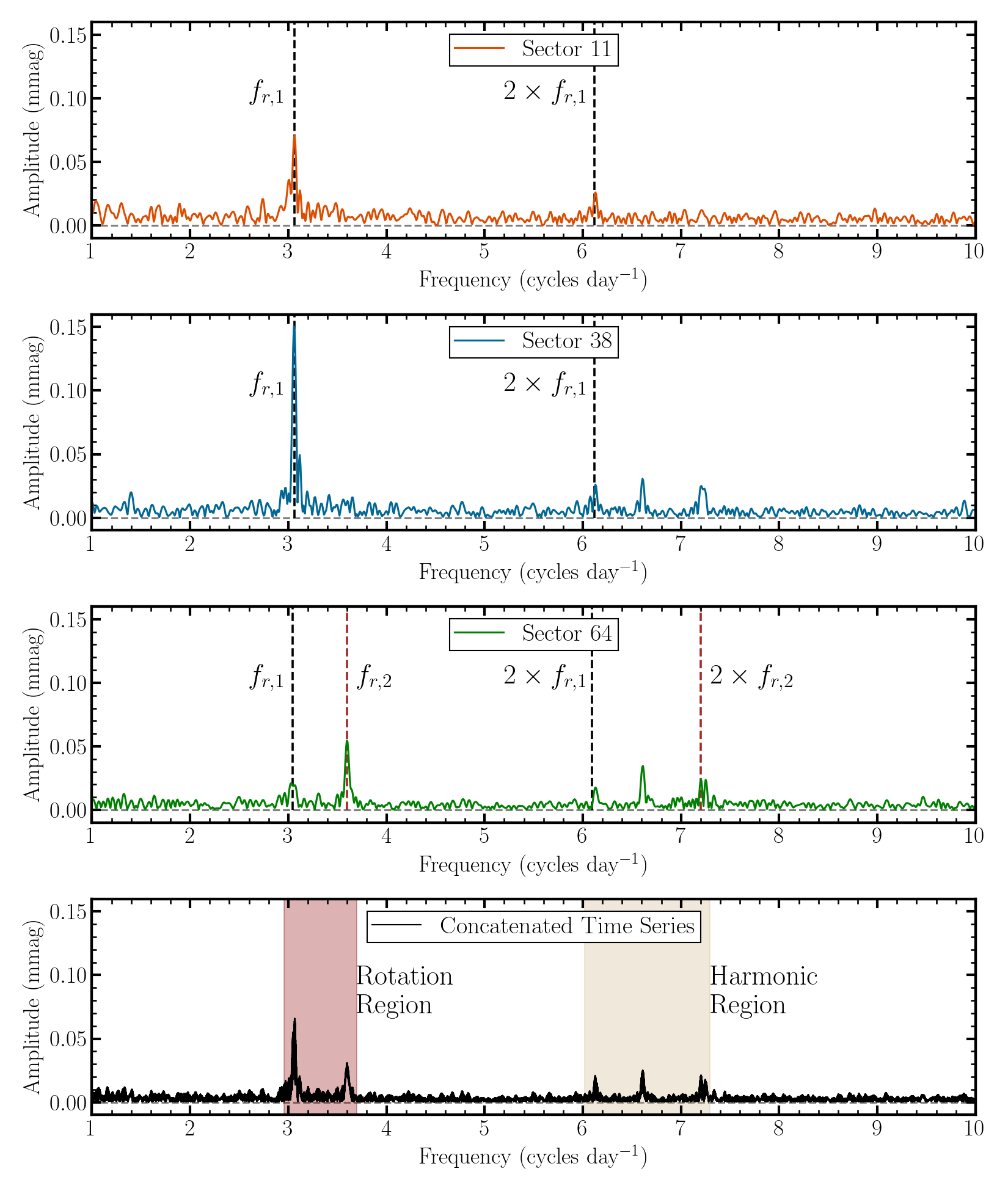}
  \centering
  \caption{TESS amplitude spectra of HIP~65426, zoomed in on a region of low frequencies. Vertical dashed lines denote the locations of $f_{r,1}$, $f_{r,2}$ as described in Section \ref{sec:rot}, as well as the locations of $2 \times f_{r,1}$ and $2 \times f_{r,2}$.  
  \label{fig:rot}} 
\end{figure*}
Inspection of the $<10$~cycles~day$^{-1}$ region of the amplitude spectrum reveals several significant signals that are well-separated from the pressure modes (Figure \ref{fig:rot}). We identify two rotation frequencies, $f_{r,1}$=3.06~cycles~day$^{-1}$ and $f_{r,2}$= 3.60~cycles~day$^{-1}$. $f_{r,1}$ is detected in all three TESS sectors whereas $f_{r,2}$ is detected is only Sector 64. 
Gaussian functions were fit to the highest peak of each frequency and the best-fitting standard deviation is the reported uncertainty (Table \ref{tab:stellarrot}). In the case of $f_{r,1}$, our adopted value is the average of the three separate detections from each TESS sector. This yielded $f_{r,1}$=3.06$\pm$0.03~cycles~day$^{-1}$ (7.85$\pm$0.08~hr) and $f_{r,2}$=3.60$\pm$0.02~cycles~day$^{-1}$ (6.67$\pm$0.04~hr). We note that $2 \times f_{r,2}$ is similar to our estimated value for $\Delta \nu$. This is likely a coincidence, but supports the conclusion that rapid rotation will complicate the mode identification for this star.

\begin{figure}[t!]
  \includegraphics[trim=0.05cm 0cm 1.4cm 1.75cm, clip,width=3.35in]{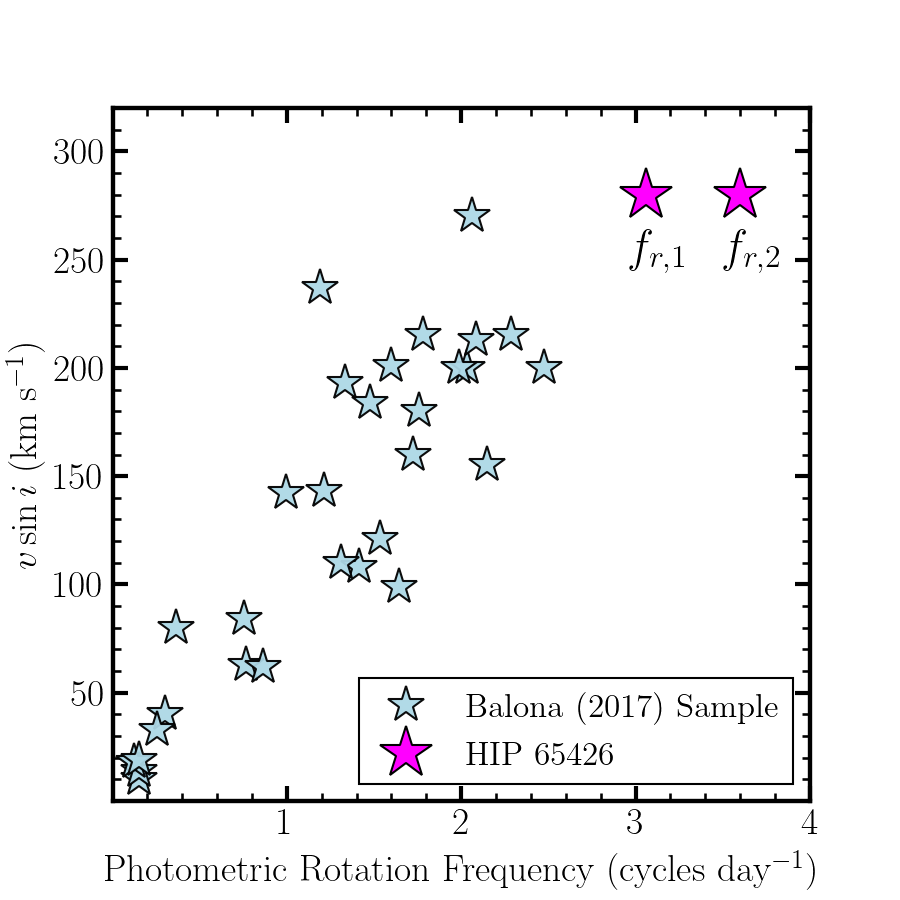}
  \centering
  \caption{$v \sin{i}$ compared to photometric rotation frequency for the sample of hot (8300--12000~K) stars presented in \citet{Balona2017}. Pink stars for HIP~65426 (8840$\pm$200~K, \citealt{Chauvin+2017}) are overplotted to represent both $f_{r,1}$ and $f_{r,2}$.  \label{fig:balona}} 
\end{figure}

We associated these signals with the stellar rotation period for the following reasons. \citet{Balona2017} compared Kepler photometric rotation frequencies for a sample of 30 hot (8300--12000~K) stars with their corresponding $v \sin{i}$ values and found that they correlated in a manner physically consistent with rotation frequencies. We overplotted $f_{r,1}$ and $f_{r,2}$ compared with the $v \sin{i}$ value of HIP~65426 (280 km~s$^{-1}$, Section \ref{sec:stellarInc}) along with the sample presented in \citet{Balona2017} in Figure \ref{fig:balona}. Both follow the expected trend, which validates our interpretation of the signals as tracers of the stellar rotation frequency. We considered if HIP~65426 could be a hybrid pulsator showing gravity modes in addition to pressure modes \citep[e.g.,][]{Grigahcene+2010}. However, low-frequency gravity modes often appear in groups of several dense peaks \citep[e.g.,][]{Li+2020}, as opposed to more narrow signals like in the case of HIP~65426. Moreover, the TESS amplitude spectra show signal at $2 \times f_{r,1}$ and $2 \times f_{r,2}$ (Figure \ref{fig:rot}) consistent with rotation harmonics and an untypical feature for pulsations \citep[e.g.,][]{Uytterhoeven+2011}. The peak at $\approx$6.6~cycles~day$^{-1}$ is associated with rotation, but we do not include it in Table \ref{tab:stellarrot} due to the absence of a subharmonic. This is not uncommon for rotational modulation \citep[e.g.][]{2021RNAAS...5..258K} and could be caused by multiple spots located at different longitudes. The same is likely true for the $2 \times f_{r,2}$ frequency in Sector 38 without $f_{r,2}$.

Together, we interpret the detection of $f_{r,1}$ and $f_{r,2}$ as evidence for surface latitudinal differential rotation in HIP~65426. The physical consistency of both frequencies with $v \sin{i}$, as well as their proximity to one another, is consistent with the two frequencies tracing different stellar latitudes rotating at different velocities. Moreover, while each TESS sector is only $\sim$27~days, they are separated by $\sim$2~years, which corroborates a picture of spot evolution occurring on those timescales. This would also explain why $f_{r,2}$ is only detected in one sector and why $f_{r,1}$ shows amplitude variations between sectors. This is consistent with \citet{Balona+Abedigamba2016}, who analyzed the photometric rotation frequency spread for a sample of 522 hot (7400--10000~K) A and F stars and found that $\approx 60 \%$ of this sample was consistent with exhibiting differential rotation. Their analysis used Kepler light curves with a continuous time baseline of $\sim$4~yr, while only three sectors of TESS data are available for HIP~65426, limiting more detailed investigation.  

We noted in Section \ref{sec:obs} an absence of nearby comparatively bright sources to HIP~65426, as well as a lack of stellar companions. We further assessed the likelihood that our reported rotation frequencies correspond to HIP~65426 and not a faint contaminating background source in the TESS aperture. To this end, we used the \texttt{TESS\_localize} package \citep{Higgins+Bell2023}, which is an algorithm designed to identify the most likely Gaia source in the TESS aperture responsible for producing a set of input frequencies. In very brief summary, the algorithm models the expected relative flux distribution of the input frequencies (i.e., the frequency amplitude) across TESS pixels to estimate the location of the variability source, assuming that only one source varies at the input frequencies. As an initial baseline, we began by running the routine to evaluate the three most significant pulsation frequencies of HIP~65426 (Table \ref{tab:freqs}), separately for Sectors 11, 38, and 64, and down to a Gaia magnitude limit of 18. The corresponding per-sector relative probabilities that the pulsation frequencies correspond to HIP~65426 are respectively 97.9\%, $>$99.9\%, and $>$99.9\%. We next ran the routine in the same manner but now to evaluate the rotation frequencies in each of the three TESS sectors (Table \ref{tab:stellarrot}). The resulting relative probabilities that these rotation frequencies correspond to HIP~65426 are respectively 88.1\%, $>$99.9\%, and $>$99.9\%. This evaluation contributes additional confidence that our TESS rotation frequencies (and pulsation frequencies) intrinsically correspond to HIP~65426.

\begin{deluxetable}{ccccc}[t]
\tablehead{
\colhead{} & \colhead{$f_{r,1}$} &\colhead{$P_{r,1}$} & \colhead{$f_{r,2}$} & \colhead{$P_{r,2}$}
\\
\colhead{} & \colhead{cycles~day$^{-1}$} & \colhead{hr}& \colhead{cycles~day$^{-1}$} & \colhead{hr}}
\caption{TESS rotation frequencies of HIP~65426. Equivalent rotation periods are included. \label{tab:stellarrot}} 
\startdata
Sector 11&3.06$\pm$0.02&7.84$\pm$0.05&-&-\\
Sector 38&3.06$\pm$0.02&7.84$\pm$0.05&-&-\\
Sector 64&3.05$\pm$0.04&7.87$\pm$0.10&3.60$\pm$0.02&6.67$\pm$0.04\\
\hline
Adopted&3.06$\pm$0.03&7.85$\pm$0.08&3.60$\pm$0.02&6.67$\pm$0.04\\
\enddata
\end{deluxetable} 

\subsection{HIP~65426 Stellar Inclination}\label{sec:stellarInc}
We used our stellar rotation period together with published measurements of $v \sin{i}$ and stellar radius to constrain the stellar inclination. These parameters are related by a geometric relation \citep[e.g.,][]{Doyle+1984,Campbell+Garrison1985}: 

\begin{equation}
\label{eqn:incl}
i_{\star} = \sin^{-1}\left( \frac{v\sin i}{v} \right) = \sin^{-1} \left( \frac{v\sin i}{2\pi R_{\star}/P_{\rm rot}} \right).
\end{equation}

\citet{Masuda+Winn2020} noted that $v \sin{i}$ is not statistically independent of the rotational velocity, $v$ (computed using $P_{\rm rot}$ and $R_{\star}$), and thus it is not appropriate to apply Monte Carlo propagation of uncertainties. They provided a Bayesian framework that properly computes $\cos{i}$ given measurements of $P_{\mathrm{rot}}, R_{\star}$, and $v \sin{i}$. \citet{Bowler+2023} expanded on this by deriving analytic expressions that approximate the framework of \citet{Masuda+Winn2020} as long as $P_{\mathrm{rot}}$ is measured to $\lesssim 20\%$. We used the \citet{Bowler+2023} expressions in this work to derive the inclination posterior of HIP~65426, $P(i_{\star} \mid P_{\mathrm{rot}}, R_{\star}, v \sin{i})$. Their expressions \citep[Eqn. 9 and 10,][]{Bowler+2023} assume uniform priors on the measured inputs, an isotropically uniform prior for $\cos{i}$, and Gaussian-distributed uncertainties on the inputs.

\citet{Chauvin+2017} reported an isochrone-derived stellar radius of $R_{\star}=1.77\pm0.05$~\Rsun, which is consistent with the TIC value of $R_{\star}=1.74\pm0.05$~\Rsun \citep{Stassun+2018,Stassun+2019} derived using the Stefan-Boltzmann law. As an input radius for HIP~65426, we used the mean of these two literature values. We note however that rapidly rotating intermediate mass stars are better described as oblate spheroids than spheres \citep[e.g.,][]{vanBelle+2001,DomicianodeSouza+2003,McAlister+2005,Monnier+2007}. Therefore, HIP~65426 is likely not well described by a single radius but rather by a radius gradient that increases from polar to equatorial latitudes. HIP~65426 is too distant (i.e., its angular size is too small) to resolve with optical/infrared interferometry, a technique that can directly constrain the polar ($R_{\rm pol}$) and equatorial ($R_{\rm eq}$) radii for nearby stars with large projected angular sizes \citep[e.g.,][]{vanBelle2012}. To account for systematic uncertainty in our input radius due to unresolved oblateness, we consider $R_{\rm pol}$ and $R_{\rm eq}$ of Altair, a nearby $\delta$~Scuti star whose surface has been interferometrically resolved and who is of similar mass and rotational velocity as HIP~65426 \citep[e.g.,][]{Buzasi+2005,Monnier+2007,Bouchaud+2020}. Using $R_{\rm pol}$ = 1.661\Rsun\ and $R_{\rm eq}$ = 2.022\Rsun\ \citep{Monnier+2007} we calculate a fractional uncerainty of $10\%$, which we adopt for our input value for HIP~65426 (i.e., $R_{\star}=1.76\pm0.18$~\Rsun.) 

For the input $v \sin{i}$, we used the mean of two reported values measured using HARPS spectroscopy: 299$\pm$9~km~s$^{-1}$ \citep{Chauvin+2017} and 261$\pm$2~km~s$^{-1}$ \citep{Petrus+2021}. For the uncertainty, we used the standard error on the mean of the HARPS values, yielding 280$\pm$13~km~s$^{-1}$. We note that in the presence of latitudinal differential rotation, the spectroscopic $v \sin{i}$ (derived without accounting for differential rotation) will represent a weighted average of the set of latitudinal velocities and not soley represent the equatorial velocity. This is supported by interferometric imaging of the nearby A-star Altair, which has a $v \sin{i}$ of 241~km~s$^{-1}$ from detailed modeling \citep{Monnier+2007} and a $v \sin{i}$ from methods that ignore differential rotation of 216~km~s$^{-1}$ \citep[mean Altair $v \sin{i}$ from:][]{Bernacca+Perinotto1970,Abt+Morrell1995,Royer+2002,Schroder+2009}. Detailed modeling of the differential rotation (and its effect on line broadening) in HIP~65426 is beyond the scope of this paper.

For the input rotation period, we took the mean of our two TESS rotation periods from Section \ref{sec:rot}. By taking the mean we appropriately match the weighted average velocity implied by the HARPS $v \sin{i}$ measurements. We opted to keep the larger of the two individual period uncertainties for our input mean period , i.e., 7.26$\pm$0.08~hr.

The resulting posterior for $P(i_{\star} \mid P_{\mathrm{rot}}, R_{\star}, v \sin{i})$ is shown in Figure \ref{fig:incs}. Here we have assumed a prograde orbit of HIP~65426~b relative to its host star, whereby the planet orbits clockwise on the sky (i.e., is inclined at $90^{\circ}<i<180^{\circ}$, Section \ref{sec:orbFit}), and thus the inclination angle parameter space spans [90,180]$^{\circ}$. The posterior median and 68\% credible interval is $i_{\star}=107_{-11}^{+12}$$^\circ$, with a maximum a posteriori of 107$^{\circ}$. This indicates that HIP~65426 is most likely oriented somewhat edge-on, consistent with a~priori expectation from the literature $v \sin{i}$ measurements. The 95\% and 99.7\% credible intervals span 91--127$^\circ$ and 90--133$^\circ$, respectively.

\subsection{HIP~65426~b Orbital Inclination}\label{sec:orbFit}
\begin{figure*}
  \includegraphics[trim=1.5cm 0cm 2.5cm 1cm, clip, width=7.3in]{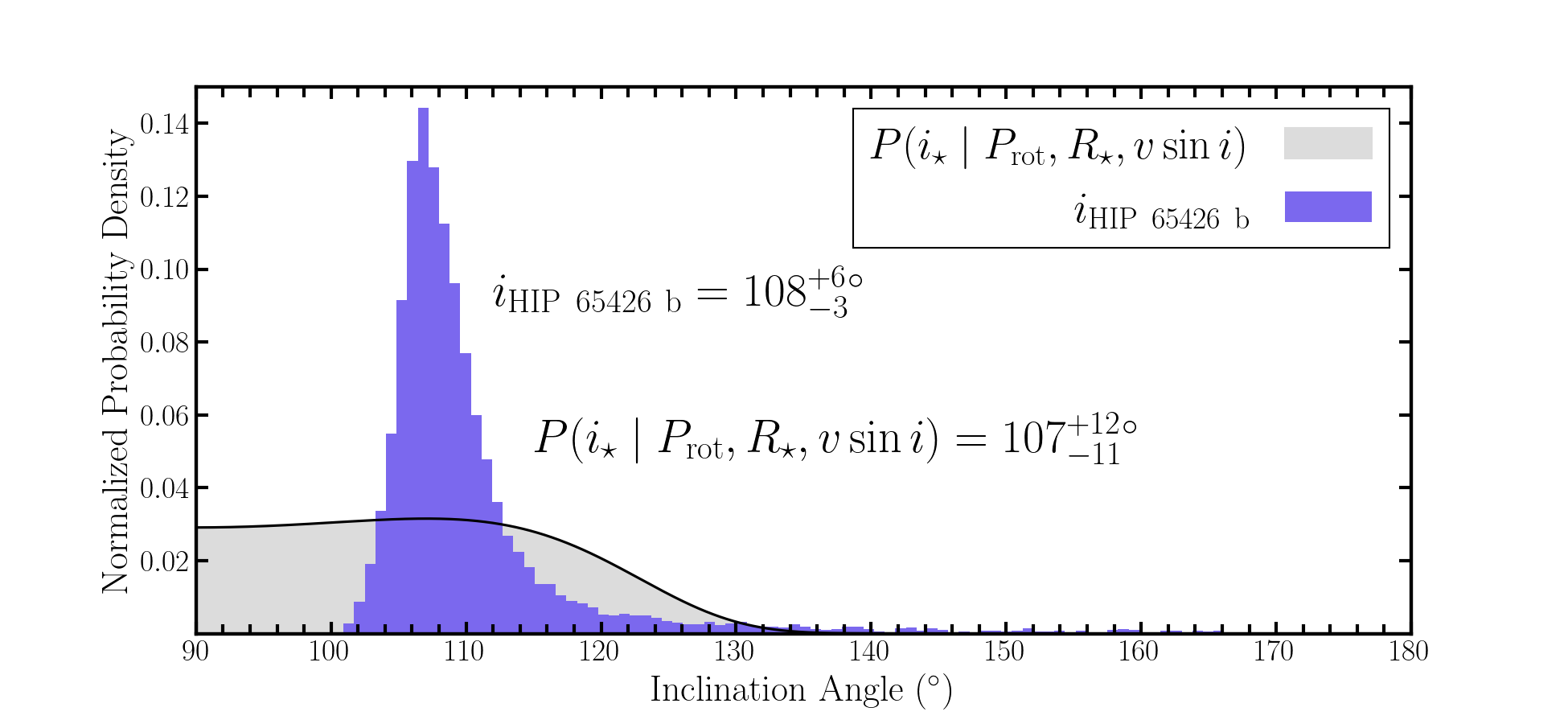}
  \centering
  \caption{Normalized inclination posteriors for the HIP~65426 system, where $P(i_{\star} \mid P_{\mathrm{rot}}, R_{\star}, v \sin{i})$ represents the host star rotational inclination and $i_{\rm HIP\ 65426\ b}$ represents the orbital inclination of HIP~65426~b. Median and 68\% credible intervals are written in the figure.
  \label{fig:incs}} 
\end{figure*}
\begin{figure*}
  \includegraphics[trim=1.5cm 0cm 2.8cm 0cm, clip, width=7.2in]{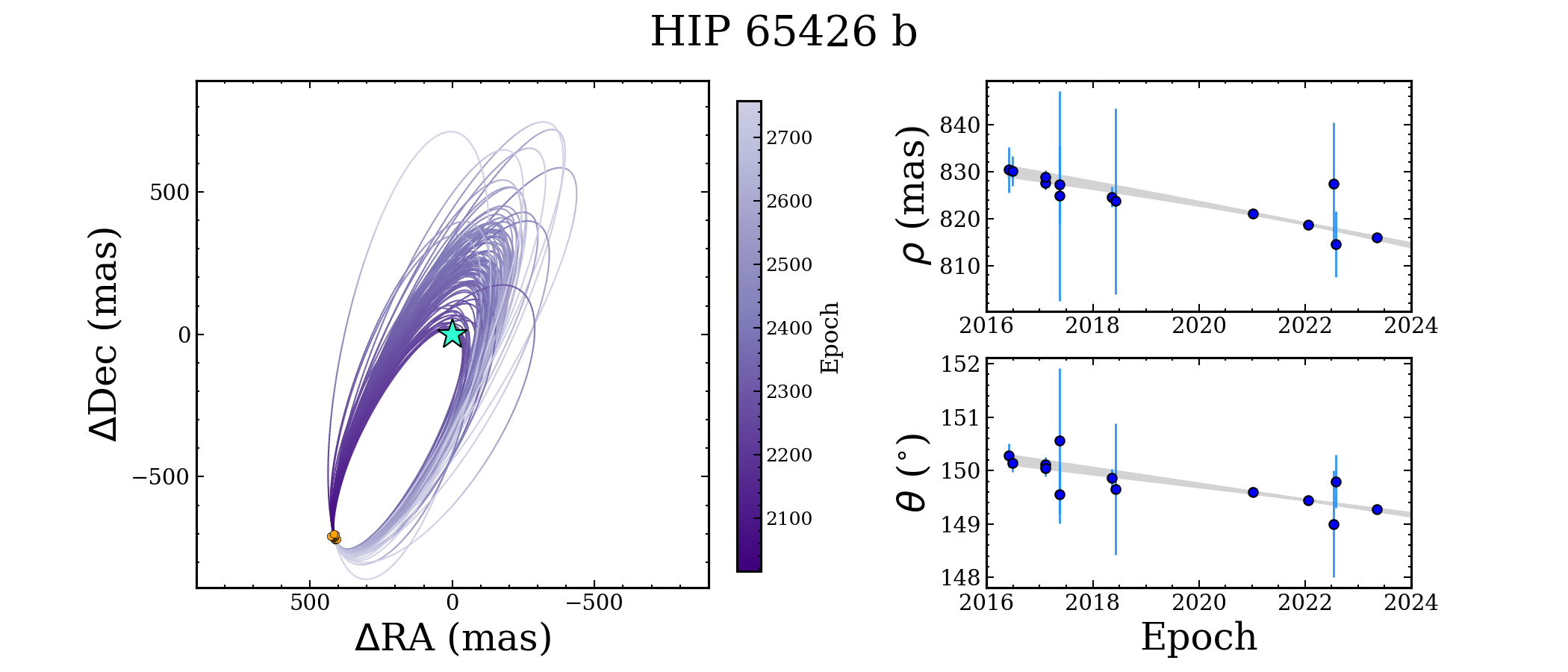}
  \centering
  \caption{The left panel displays a sample of 100 sky-projected posterior orbits resulting from our fit of HIP~65426~b. The cyan star represents the position of HIP~65426 and the orange dots represent the relative astrometry of HIP~65426~b (with no error bars on this panel). The right panels compare the model orbit separation and position angles from the same 100 posterior orbits to the input astrometric data. 
  \label{fig:orbit}} 
\end{figure*}
We conducted an orbit fit utilizing all the relative astrometry of HIP~65426~b available in the literature \citep{Chauvin+2017,Cheetham+2019,Stolker+2020,Carter+2023}, which includes three recently reported high-precision measurements with VLTI/GRAVITY \citep{Blunt+2023}. Where multiple astrometric measurements from the same instrument and same night of observing are reported, we took their mean and treated them as a single epoch as follows. From \citet{Cheetham+2019}, we averaged the two SPHERE 2018.36 epochs into $[\rho,\theta] = [824.65\pm 2.2~\rm{mas}, 149.87\pm 0.16^\circ]$. From \citet{Carter+2023}, we averaged the two reported astrometric measurements from MIRI (2022.54) as well as five from NIRCam (2022.57) into two epochs, respectively $[\rho,\theta] = [829.5\pm 13~\rm{mas}, 149\pm 1^\circ]$ and $[\rho,\theta] = [819\pm 6.2~\rm{mas}, 149.84\pm 0.42^\circ]$. \citet{Stolker+2020} reprocessed two NaCo epochs originally reported in \citet{Cheetham+2019}. We chose to use the values reported in \citet{Stolker+2020}, specifically using their final reported measurements after they applied their astrometric bias corrections.  This in total amounts to 13 astrometric epochs spanning a $\sim$7~year baseline (2016.41--2023.35). 

We considered inflating the astrometric uncertainties to account for potential systematics following the  approach of some recent orbit fitting studies with relative astrometry \citep{Bowler+2020,Sepulveda+Bowler2022}. In short, a linear model is assumed for $\rho$ and $\theta$ as a function of time, and a reduced $\chi^2$ ($\chi_{\nu} ^2$) is calculated. For any astrometric measurements only reported in [$\Delta$RA,$\Delta$Dec] form, we converted to $[\rho,\theta]$ in a Monte Carlo fashion. If $\chi_{\nu} ^2$ was greater than unity, we iteratively increased a ``jitter" uncertainty term added in quadrature to the base uncertainty until $\chi_{\nu} ^2$ reaches unity. However, for the HIP~65426~b dataset used in this study, $\chi_{\nu} ^2$ was already less than unity for both $\rho$ and $\theta$ (respectively 0.4 and 0.3, with $\nu$=11), which suggests that the uncertainties are likely reasonable and we thus incorporated no additional uncertainty. A summary figure is displayed as Figure \ref{fig:astrometry}. 

We fitted Keplerian orbits to our astrometric dataset using \texttt{orbitize!} \citep{orbitize}. We used \texttt{ptemcee} mode \citep{emcee,ptemcee}, which utilizes the parallel-tempered affine-invariant ensemble sampler of Markov chain Monte Carlo \citep{Goodman+Weare2010}. The varied parameters are standard for a Keplerian orbit with relative astrometry: $a$ (semi-major axis), $e$ (eccentricity), $i$ (inclination), $\omega$ (argument of periastron of the planet's orbit), $\Omega$ (position angle of the ascending node), $\tau$ (time of last periapsis, as defined in \citealt{orbitize} with $t_{\rm ref }$=MJD~58849), $\varpi$ (parallax), and $M_{\rm tot}$ (total mass of the system). We chose a Gaussian $M_{\rm tot}$ prior of 2.0$\pm$0.1 \Msun, consistent with previous mass estimates of HIP~65426 \citep[][]{Tetzlaff+2011,Chauvin+2017,Bochanski+2018}. The $\varpi$ prior was a Gaussian of 9.303$\pm$0.035~mas from Gaia DR3 \citep{gaiamission,GaiaDR3}. The prior for $a$ was log-uniform ranging from 1--500 au. For $i$ we used an isotropic ($P(i)\propto \sin i$) prior ranging from 0--$\pi$ rad. We use linearly uniform priors for the remaining parameters with ranges: 0--1 for $e$; 0--2$\pi$ rad for $\omega$ and $\Omega$; and 0--1 for $\tau$. To assist convergence, we chose to initialize the walkers at positions near the best-fit values as guided from preliminary orbit fits that we conducted as well as from prior orbit fits in the literature \citep[e.g.,][]{Carter+2023,Blunt+2023}. We ran \texttt{ptemcee} with 1500 walkers, 18 temperatures, and for $1.6\e{5}$ steps per walker per temperature. A burn-in size of the first 50\% of steps are removed, and we also apply a thinning factor of 20 to mitigate the effect of correlation. Keeping only the samples of the lowest temperature yields a final total of $6\e{6}$ posterior orbit samples. A sample of sky-projected posterior orbits are displayed as Figure \ref{fig:orbit}, and a selection of posterior orbital elements are displayed in Appendix \ref{appendix:orbitFit}.

The most important result for the purposes of our study is the orbital inclination posterior ($i_{\rm HIP\ 65426\ b}$), which is shown in Figure \ref{fig:incs}. The median and 68\% credible interval is $108_{-3}^{+6}$$^{\circ}$; the 95\% credible interval spans 103--139$^{\circ}$. Our inclination posterior is consistent within the 68\% uncertainties to those of recent studies \citep[e.g.,][]{Carter+2023,DoO+2023,Blunt+2023}, despite minor differences in the choices of input astrometry and priors. 

\section{Discussion}\label{secdiscussion}
\subsection{A Lack of Evidence for Misalignment}
The inclination posteriors for HIP~65426~b and its host star are compared in Figure \ref{fig:incs}. They are consistent within their 68\% uncertainties, which we interpret as a lack of significant evidence for misalignment. In the absence of knowledge of the stellar rotation axis orientation, the best available proxy for the true star-planet obliquity is the inclination difference of HIP~65426~b and its host star, $|\Delta i| = |i_{\star}-i_{\rm HIP\ 65426\ b}|$, which represents a minimum inclination difference with respect to the true obliquity. We place 68\% and 95\% upper limits of $|\Delta i| < 13$$^{\circ}$ and $|\Delta i| < 25$$^{\circ}$, respectively, for the HIP~65426 system.

\subsection{Formation of HIP~65426~b}
The wide orbit of HIP~65426~b is particularly interesting with respect to its formation history because core accretion is not expected to operate efficiently at separations of several tens of au \citep[e.g.,][]{Pollack+1996,DodsonRobinson+2009,Rafikov2011}. Coupled with the large eccentricity implied by early orbit fitting studies, planet-planet scattering has been proposed as a possible contributing mechanism for HIP~65426~b's presently-observed orbit. \citet{Marleau+2019} explored planet-planet scattering scenarios in detail using the system properties together with $N$-body modeling. If HIP~65426~b formed via core accretion at a location interior to its present-day orbit, \citet{Marleau+2019} suggest the presence of additional yet-undetected interior planets that would have participated in gravitationally scattering HIP~65426~b. Our lack of significant evidence for misalignment stands to disfavor a core accretion followed-by planet-planet scattering scenario because the latter is generally expected to result in high mutual inclinations \citep{Chatterjee+2008,Ford+Rasio2008}. Although with incomplete geometric information, it is still possible that the true obliquity of HIP~65426 could be larger than $|\Delta i|$.

It is important to note that the orbital eccentricity of HIP~65426~b, which has played a key role in preliminary interpretations of its history, is not yet concretely determined. The most recent orbit fit by \citet{Blunt+2023} using high-precision astrometry from VLTI/GRAVITY (as well as this work, which incorporated their GRAVITY data) disfavor the largest orbital eccentricities that were previously plausible based on eccentricity posteriors from earlier studies. \citet{Blunt+2023} also showed that even with three GRAVITY epochs, the choice of eccentricity prior still influences the posterior. \citet{DoO+2023} previously demonstrated this as well, where their adoption of an observable-based eccentricity prior \citep{ONeil+2019} yielded an eccentricity posterior for HIP~65426~b more consistent with lower eccentricities compared to that derived with a uniform eccentricity prior. Continued astrometric monitoring to determine an accurate orbital eccentricity that is prior-independent will be needed to better understand the formation of HIP~65426~b. 

\subsection{Trends for Directly-Imaged Giant Planets}\label{subsec:DITrends}
\begin{figure}[t!]
  \includegraphics[trim=0cm 0cm 1cm 1cm, clip, width=3.35in]{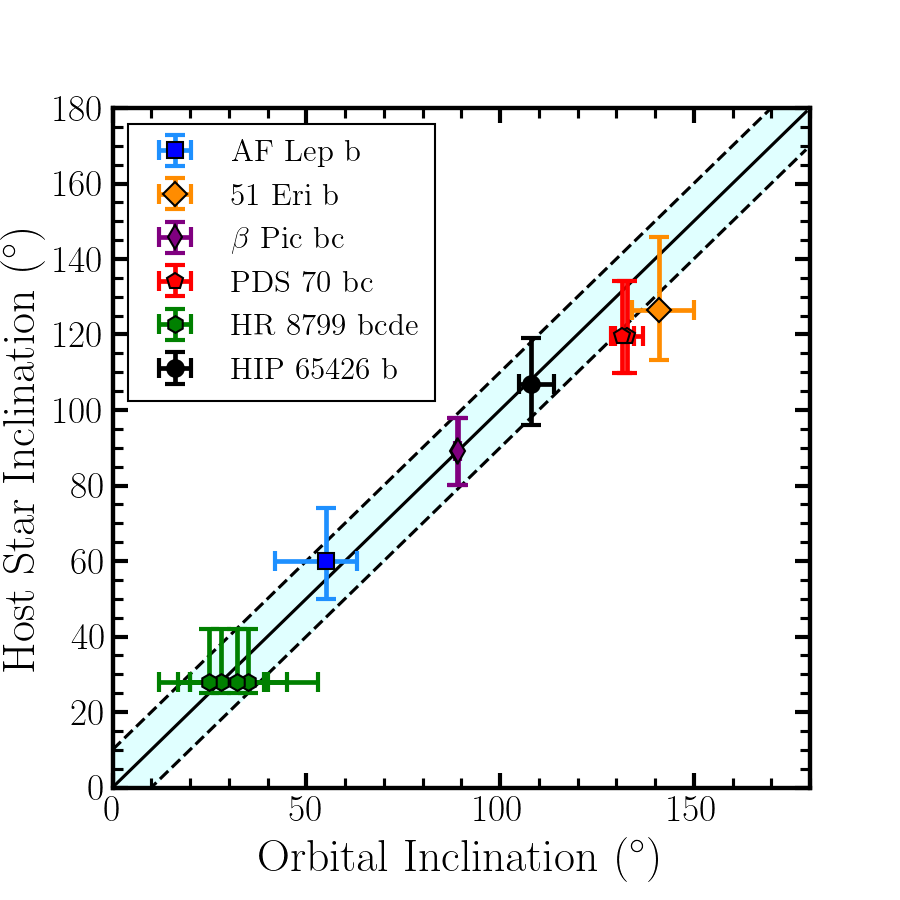}
  \centering
  \caption{Orbital inclination compared to stellar inclination for a sample of 6 directly imaged exoplanet systems (comprising 11 total planets) assuming prograde orbits with respect to the host star. The solid black line denotes the 1:1 relation, and the light cyan region bordered by dashed black lines encapsulates an inclination difference $\pm$10$^{\circ}$ of the black line. References for the inclination angles are given in Section \ref{subsec:DITrends}. \label{fig:GPTrends}} 
\end{figure}
The lack of evidence for misalignment for HIP~65426~b furthers an emerging trend where imaged long-period giant planets appear preferentially aligned with their host stars \citep{Bowler+2023}. We illustrate this in Figure \ref{fig:GPTrends} where orbital inclinations and host star inclinations are compared for 6 directly imaged exoplanet systems comprising 11 total companions: HR~8799~bcde \citep{marios2008,marios2010}, 51~Eri~b \citep{Macintosh+2015,DeRosa+2015}, PDS~70~bc \citep{Keppler+2018,Haffert+2019,Mesa+2019}, $\beta$~Pic~bc \citep{Lagrange+2009,Lagrange+2010,Lagrange+2019,Nowak+2020}, AF~Lep~b \citep{Franson+2023,Mesa+2023,DeRosa+2023}, and HIP~65426~b \citep{Chauvin+2017}. The stellar and orbital inclination references are as follows: HR~8799 \citep{Sepulveda+Bowler2022,Sepulveda+2023}, 51~Eri \citep{Dupuy+2022,Bowler+2023}, PDS~70 \citep{Wang+2021,Bowler+2023}, $\beta$~Pic \citep{Zwintz+2019,Brandt+2021}, AF~Lep \citep{Zhang+2023}, and HIP~65426 (this work). We use median values (or alternatively maximum a posteriori values if no median is reported) and 68\% uncertainties to represent the error bars. For $\beta$~Pic\footnote{$\beta$~Pictoris is a rare exception among directly-imaged exoplanet systems where the full obliquity has been constrained. The proximity of $\beta$~Pictoris enabled \citet{Kraus+2020} to interferometrically determine the sky-projected position angle of its equator. Together with the asteroseismic stellar inclination from \citet{Zwintz+2019}, \citet{Kraus+2020} determined a true mutual inclination angle of $\leq3\pm 5^{\circ}$ for $\beta$~Pic~b and its host star.}, no uncertainty is reported on the asteroseismic stellar inclination estimate from \citet{Zwintz+2019}, and we thus adopted a 10\% uncertainty. The imaged giant planet sample in Figure \ref{fig:GPTrends} are all unambiguously consistent with $|\Delta i| <10^{\circ}$, provided the adopted inclination uncertainties.

Dusty debris disks, which represent extrasolar Kuiper Belt analogs \citep[e.g.,][and references therein]{Hughes+2018,Marino2022}, were also shown to be preferentially aligned with their host stars based on early studies with small (N=8--10) sample sizes \citep{Watson+2011,Greaves+2014}. Recently, \citet{Hurt+MacGregor2023} evaluated $|\Delta i|$ for a larger sample size of 31 debris disk systems. While they found some examples of significantly misaligned systems, the lack of significant evidence for misalignment persisted in $\gtrsim$80\% of their sample, which still suggests that most debris disks are preferentially aligned with their host stars. Additional observations will show whether the link in the formation and dynamical evolution of imaged giant planets and debris disks continues to hold for obliquities.

If this trend does hold true for imaged giant planets, it would contrast with the emerging obliquity trends found for imaged brown dwarfs, which imply that misalignments are common \citep{Bowler+2023}. Together with distinctions in other fundamental properties (e.g., underlying orbital eccentricity distributions, \citealt{Bowler+2020,Nagpal+2023,DoO+2023}) this could be further illustrating brown dwarfs and giant planets as distinct classes of objects with differing formation pathways. Future direct imaging discoveries that can include a TESS analysis of the host star will help to create a larger and more diverse sample size to determine whether this population-level obliquity trend for imaged giant planets holds true.

\section{Conclusion}
We analyzed TESS time-series photometry of the directly-imaged exoplanet host star HIP~65426. We detected several pulsation modes consistent with classification as a high-frequency $\delta$~Scuti pulsator. We find a preliminary estimate of $\Delta \nu =$7.23$\pm$0.02~cycles~day$^{-1}$, consistent with other young $\delta$~Scuti stars \citep{Bedding+2020Natur}. We also used the nondetection of pulsation timing variations to estimate a 95\% dynamical mass upper limit of 12.8~\Mjup\ for HIP~65426~b, independently supporting its planetary mass.

From the TESS data we also detected rapid photometric rotation frequencies of 7.85$\pm$0.08~hr and 6.67$\pm$0.04~hr that we interpreted as evidence of latitudinal differential rotation in HIP~65426. Using published radius and $v \sin{i}$ constraints together with our TESS rotation frequencies, we jointly measured the inclination of HIP~65426 to $i_{\star}=107_{-11}^{+12}$$^\circ$. This near edge-on inclination of the host star is consistent with HIP~65426~b's near edge-on inclination of  $108_{-3}^{+6}$$^{\circ}$ that we determined from our orbit fit of the available relative astrometry. We thus found no significant evidence for spin-orbit misalignment in the HIP~65426 system. This finding further supports an emerging trend of preferential alignment between imaged long-period giant planets and their host stars (Figure \ref{fig:GPTrends}), which is similar to the emerging trend for debris disks and dissimilar to that for imaged brown dwarfs.

This work adds to the synergy of using space-based photometric data and direct imaging data to probe the \emph{outer} architectures of extrasolar systems at an epoch after planet formation has occurred. Evaluating the time-series photometry of additional host stars of imaged exoplanet, brown dwarf, and debris disk systems will be key to investigating the emerging obliquity trends. Likewise, dedicated missions and surveys to discover new systems and robustly characterize the inclinations of their constituents will also be a complementary necessity.

\textit{Acknowledgments.}
AGS thanks Brendan P. Bowler and Sarah Blunt for collaborative discussions that helped expedite this work, as well as Mitchell T. Dennis for assistance with operating the HPC clusters Mana and Koa.

This material is based upon work supported by the National Science Foundation Graduate Research Fellowship Program under Grant No. 1842402 and 2236415. D.H. acknowledges support from the Alfred P. Sloan Foundation, the National Aeronautics and Space Administration (80NSSC21K0784), and the Australian Research Council (FT200100871).

This work has benefitted from The UltracoolSheet \citep{UCSZenodo}, maintained by Will Best, Trent Dupuy, Michael Liu, Rob Siverd, and Zhoujian Zhang, and developed from compilations by \citet{Dupuy+Liu2012}, \citet{Dupuy+Kraus2013}, \citet{Liu+2016}, \citet{Best+2018}, and \citet{Best+2021}. 

The technical support and advanced computing resources from University of Hawaii Information Technology Services – Cyberinfrastructure, funded in part by the National Science Foundation MRI award \# 1920304, are gratefully acknowledged.

This paper includes data collected by the TESS mission, which are publicly available from the Mikulski Archive for Space Telescopes (MAST). Funding for the TESS mission is provided by the NASA's Science Mission Directorate. 

This work has made use of data from the European Space Agency (ESA) mission {\it Gaia} (\url{https://www.cosmos.esa.int/gaia}), processed by the {\it Gaia} Data Processing and Analysis Consortium (DPAC, \url{https://www.cosmos.esa.int/web/gaia/dpac/consortium}). Funding for the DPAC has been provided by national institutions, in particular the institutions participating in the {\it Gaia} Multilateral Agreement. 

This research has made use of the VizieR catalogue access tool, CDS, Strasbourg, France (DOI : 10.26093/cds/vizier). The original description of the VizieR service was published in \citet{Ochsenbein+2000}.

This research has made use of the SIMBAD database, operated at CDS, Strasbourg, France. 

This research has made use of NASA's Astrophysics Data System Bibliographic Services.

\facilities{TESS}

\software{\texttt{lightkurve} \citep{LightkurveCollaboration+2018}, \texttt{SigSpec} \citep{SigSpec}, \texttt{echelle} \citep{Hey+Ball2020}, \texttt{orbitize!} \citep{orbitize}, \texttt{ptemcee} \citep{emcee,ptemcee}, \texttt{TESS\_localize} \citep{Higgins+Bell2023}, \texttt{matplotlib} \citep{matplotlib},  \texttt{astropy} \citep{astropy,astropy2018}, \texttt{numpy} \citep{Harris+2020}, \texttt{scipy} \citep{scipy}, \texttt{astroquery} \citep{astroquery}}

\appendix
\section{HIP~65426 Pulsation Frequencies}
Here we tabulate the HIP~65426 pulsation frequencies that we extracted from each sector of TESS photometry (Table \ref{tab:freqs}). We report the cumulative spectral significance and Fourier-domain phase angles as defined in \citet{Reegen+2011}. Uncertainties were calculated following \citet{Kallinger+2008}. 

\begin{deluxetable*}{cccc}
\tablehead{
\colhead{Frequency} & \colhead{Amplitude} & \colhead{Phase Angle} & \colhead{Spectral Significance}\\
\colhead{(cycles~day$^{-1}$)} & \colhead{(mmag)} &\colhead{(rads)}&\colhead{}}
\caption{Significant pulsation frequencies of HIP~65426 from TESS photometry.\label{tab:freqs}} 
\startdata
\cutinhead{Sector 11} 
27.547$\pm$0.006&0.059$\pm$0.009&$-$2.58$\pm$0.07&46\\
37.029$\pm$0.005&0.066$\pm$0.009&$-$2.15$\pm$0.06&56\\
38.499$\pm$0.007&0.045$\pm$0.009&$-$0.73$\pm$0.09&28\\
43.995$\pm$0.007&0.050$\pm$0.009&$-$2.64$\pm$0.08&33\\
49.664$\pm$0.007&0.045$\pm$0.009&0.03$\pm$0.09&28\\
53.060$\pm$0.004&0.089$\pm$0.009&$-$1.43$\pm$0.05&98\\ 
55.800$\pm$0.003&0.114$\pm$0.009&1.62$\pm$0.04&153\\  
64.821$\pm$0.003&0.129$\pm$0.009&1.64$\pm$0.03&185\\ 
66.363$\pm$0.004&0.077$\pm$0.009&$-$0.32$\pm$0.05&75\\
73.606$\pm$0.007&0.047$\pm$0.009&0.82$\pm$0.08&30\\
77.21$\pm$0.01&0.034$\pm$0.009&2.7$\pm$0.1&16\\ 
91.56$\pm$0.01&0.028$\pm$0.008&1.0$\pm$0.1&11\\
130.83$\pm$0.01&0.029$\pm$0.009&0.3$\pm$0.1&12\\
\cutinhead{Sector 38} 
27.550$\pm$0.005&0.054$\pm$0.008&$-$2.63$\pm$0.07&50\\
37.028$\pm$0.008&0.036$\pm$0.007&$-$0.4$\pm$0.1&23\\
38.502$\pm$0.005&0.053$\pm$0.008&$-$2.17$\pm$0.07&49\\
43.996$\pm$0.005&0.060$\pm$0.008&$-$0.57$\pm$0.06&60\\
53.059$\pm$0.006&0.044$\pm$0.008&0.11$\pm$0.08&34\\
53.26$\pm$0.01&0.028$\pm$0.007&$-$1.0$\pm$0.1&14\\
55.801$\pm$0.007&0.038$\pm$0.008&0.33$\pm$0.09&26\\
57.550$\pm$0.008&0.034$\pm$0.007&1.5$\pm$0.1&21\\
61.182$\pm$0.008&0.036$\pm$0.008&3.1$\pm$0.1&23\\
64.82$\pm$0.01&0.027$\pm$0.007&1.0$\pm$0.1&13\\
66.363$\pm$0.003&0.090$\pm$0.008&$-$1.45$\pm$0.04&130\\ 
73.604$\pm$0.003&0.116$\pm$0.008&0.38$\pm$0.03&206\\  
80.814$\pm$0.005&0.059$\pm$0.008&2.25$\pm$0.06&58\\ 
130.84$\pm$0.01&0.024$\pm$0.007&$-$0.6$\pm$0.1&11\\
\cutinhead{Sector 64} 
27.550$\pm$0.006&0.036$\pm$0.006&0.35$\pm$0.08&36\\ 
38.507$\pm$0.006&0.036$\pm$0.006&2.10$\pm$0.08&38\\ 
44.001$\pm$0.006&0.038$\pm$0.006&1.20$\pm$0.07&39\\ 
49.663$\pm$0.007&0.034$\pm$0.006&$-$1.23$\pm$0.08&32\\ 
53.052$\pm$0.006&0.054$\pm$0.008&1.02$\pm$0.07&45\\ 
53.266$\pm$0.007&0.028$\pm$0.005&2.67$\pm$0.09&26\\ 
55.793$\pm$0.006&0.044$\pm$0.007&0.33$\pm$0.07&42\\ 
57.549$\pm$0.005&0.053$\pm$0.008&$-$3.10$\pm$0.07&45\\ 
61.18$\pm$0.01&0.027$\pm$0.007&$-$0.4$\pm$0.1&17\\ 
64.821$\pm$0.003&0.090$\pm$0.006&$-$2.56$\pm$0.03&193\\ 
66.363$\pm$0.002&0.116$\pm$0.007&1.92$\pm$0.03&257\\
73.604$\pm$0.004&0.059$\pm$0.006&$-$2.66$\pm$0.05&105\\ 
77.21$\pm$0.01&0.024$\pm$0.006&1.8$\pm$0.1&15\\ 
80.811$\pm$0.003&0.060$\pm$0.005&$-$1.45$\pm$0.04&139\\ 
\enddata
\end{deluxetable*}

\newpage
\section{HIP~65426~b Orbit Fit}\label{appendix:orbitFit}
\begin{figure*}
  \includegraphics[trim=0cm 0cm 0cm 0cm, clip, width=6.365in]{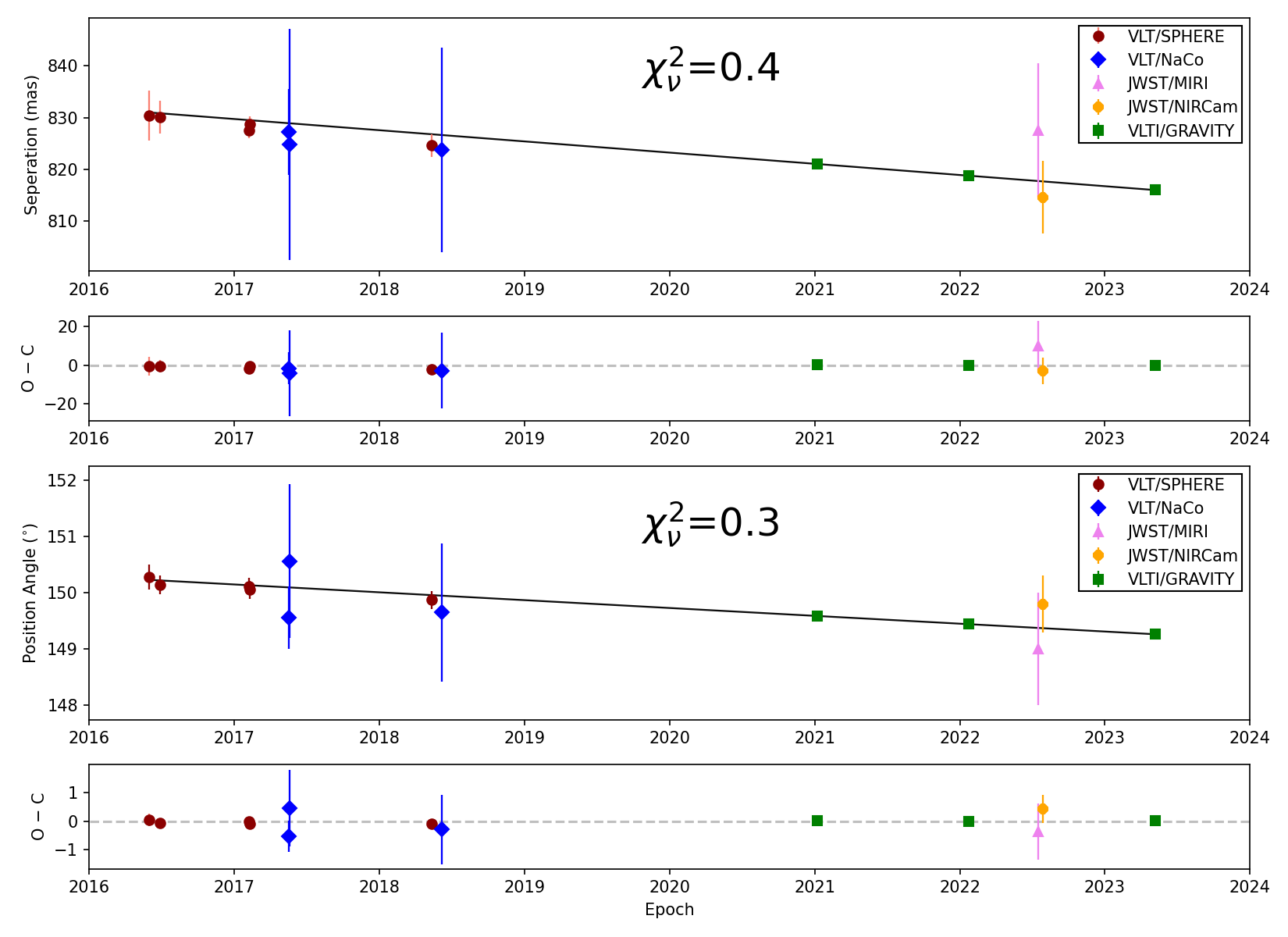}
  \centering
  \caption{Relative astrometry of HIP~65426~b used in our orbit fit (Section \ref{sec:orbFit}) compared to our best-fitting linear models used to evaluate the astrometric uncertainties. Even with the inclusion of high-precision measurements from GRAVITY, the uncertainties overall appear reasonable.
  \label{fig:astrometry}} 
\end{figure*}
\begin{figure*}
  \includegraphics[trim=0cm 0cm 0cm 0cm, clip, width=6.365in]{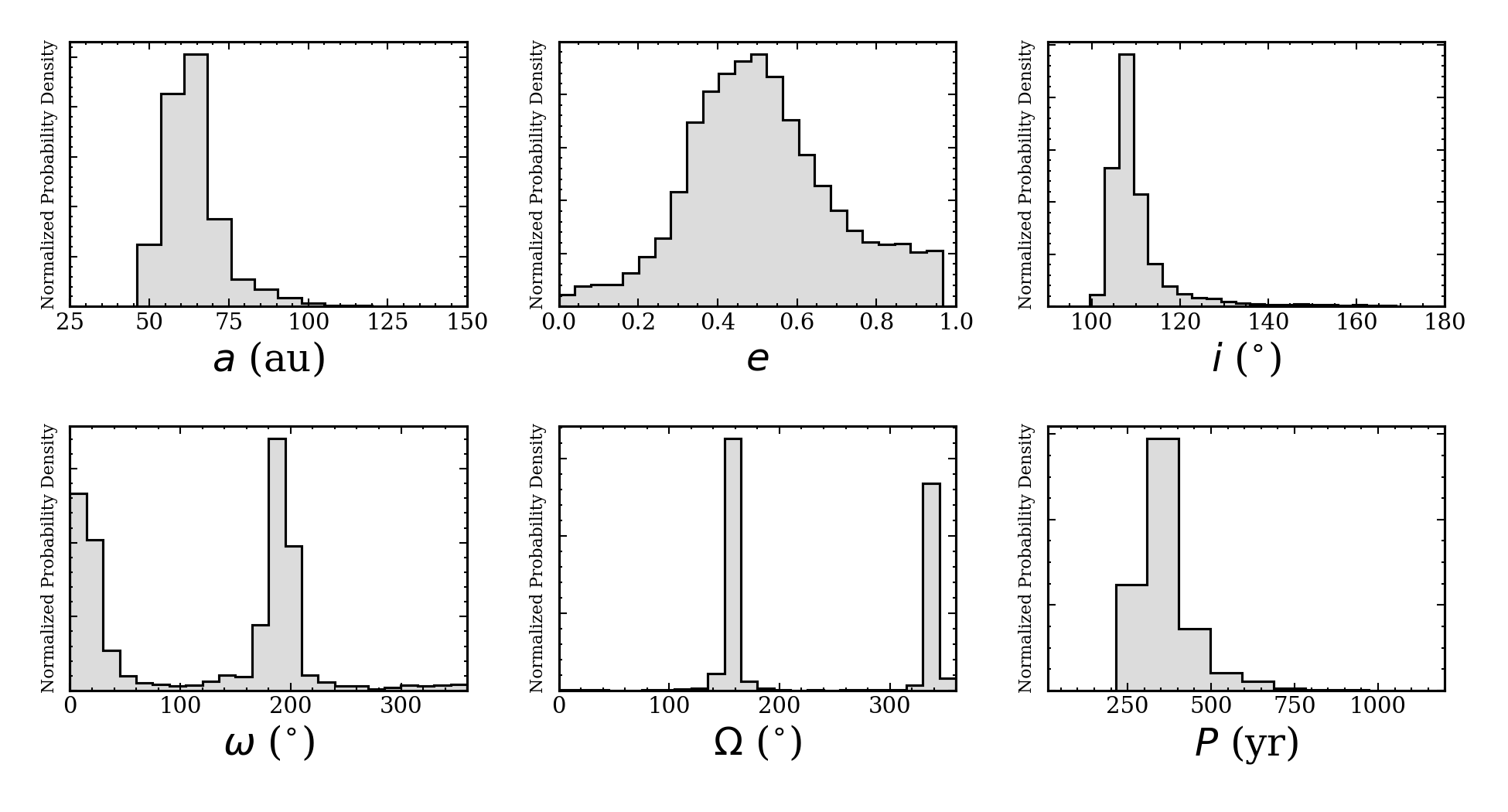}
  \centering
  \caption{1D posterior probability densities from our orbit fit of HIP~65426~b.
  \label{fig:orbitElements}} 
\end{figure*}

Here we display the relative astrometry used in our orbit fit along with the best-fitting linear models used to evaluate the astrometric uncertainties (Figure \ref{fig:astrometry}). The 1D marginalized posterior distribution for a selection of orbital parameters are displayed in Figure \ref{fig:orbitElements}. The orbital period ($P$) posterior is generated using our $a$ and $M_{\rm tot}$ posterior samples and applying Kepler's 3rd Law in a monte carlo fashion. We note that our posterior for $M_{\rm tot}$, which is not displayed in Figure \ref{fig:orbitElements}, is unconstrained by the astrometric data and merely recovers our prior probability choice.

\clearpage
\bibliographystyle{aasjournal}
\bibliography{ms}

\end{document}